\documentclass[11pt]{article}
\usepackage{verbatim,vmargin,graphicx,amsmath,calc,epsf,epsfig,subfigure,rotating}

\newcommand{\BABARPubYear}    {04}

\newcommand{\BABARConfNumber} {012}
\newcommand{\SLACPubNumber} {10639}

\input ./pubboard/babarsym

\def\Dcp    {\ensuremath{D^{0}_{\CP }}\xspace}
\def\Acp    {\ensuremath{{\cal A}_{\CP }}\xspace}
\def\Acpp   {\ensuremath{{\cal A}_{\CP+ }}\xspace}
\def\Acpm   {\ensuremath{{\cal A}_{\CP- }}\xspace}
\def\Acppm  {\ensuremath{{\cal A}_{\CP \pm }}\xspace}
\def\Rcp    {\ensuremath{{\cal R}_{\CP }}\xspace}
\def\Rcpp   {\ensuremath{{\cal R}_{\CP+ }}\xspace}
\def\Rcpm   {\ensuremath{{\cal R}_{\CP- }}\xspace}
\def\Rcppm  {\ensuremath{{\cal R}_{\CP \pm }}\xspace}
\def\cpp    {\ensuremath{\CP+ }\xspace}
\def\cpm    {\ensuremath{\CP- }\xspace}
\def\cppm   {\ensuremath{\CP \pm }\xspace}

\def\dzcppm {\ensuremath{{\Dz}_{\cppm }}\xspace}
\def\de {\ensuremath{{\rm \Delta}E}\xspace}
\def\pdf {\ensuremath{{PDF}}\xspace}

\newif\ifdimspec
\def\figsize#1{\dimspecfalse \checkdim#1\end
\ifdimspec
  \def\figureWidth{#1}%
\else
  \def\figureWidth{#1 in}\fi}
\def\checkdim#1{\ifx#1\end \let\next=\relax
  \else \ifcat#1a \dimspectrue \fi \let\next=\checkdim\fi \next}
\newcommand{\lblcaption  }[2]{\caption{#2\label{fig:#1}}}
\newcommand{\Figure }[3]{\begin{figure}[!ht]
  \begin{center}
  \mbox{\figsize{#3}\epsfig{file=#1.eps,width=\figureWidth}}
  \end{center}
  \lblcaption{#1}{#2}
  \end{figure}}
\newcommand{\twoAngFiguresEPS}[6]
{
\begin{figure}
\begin{center}
\figsize{#4}
\begin{minipage}[t]{3.2in}
\begin{center}
\epsfig{file=#1.eps,width=\figureWidth,angle=#5}
\end{center}
\end{minipage}
\begin{minipage}[t]{3.2in}
\begin{center}
\epsfig{file=#2.eps,width=\figureWidth,angle=#6}
\end{minipage}
\end{center}
\lblcaption{#1}{#3}
\end{figure}
}

\long\def\inst#1{\par\nobreak\kern 4pt\nobreak
    {\it #1}\par\vskip 10pt plus 3pt minus 3pt}

\setpapersize{USletter}
\setmarginsrb{1.0in}{1.5in}{1.0in}{0.8in}%
{0.0\baselineskip}{0.0\baselineskip}{0.0\baselineskip}{2.0\baselineskip}

\begin{document}
{\pagestyle{empty}


\vspace*{-1.5cm}

\begin{flushright}
\babar-CONF-\BABARPubYear/\BABARConfNumber \\
SLAC-PUB-\SLACPubNumber \\
August 2004 \\
\end{flushright}

\par\vskip 3cm

\begin{center}
\Large \bf Measurement of \boldmath{\CP}-Asymmetries for the Decays \boldmath{\Bpm \to \Dcp \Kstarpm}  with the BABAR Detector
\end{center}
\bigskip

\begin{center}
\large The \babar\ Collaboration\\
\mbox{ }\\
\today
\end{center}
\bigskip

\begin{center}
\large \bf Abstract
\end{center}
Using a sample of 227 million \FourS \to \BB events collected with the
\babar\ detector at the \pep2 \BF\ in 1999--2004, 
we study \Bm \to \Dz \Kstar(892)$^-$ decays where \Kstarm \to \KS \pim\ and
\Dz\ \to  \Km\pip, \Km\pip\piz, \Km\pip\pip\pim (non-\CP final states), 
\KpKm, \pip\pim (\cpp\ eigenstates), \KS~\piz, \KS$\phi$ and \KS$\omega$ (\cpm eigenstates). 
The partial rate charge asymmetries \Acp\ and the ratios \Rcp\ defined in the literature as the sum of 
the \Bp and \Bm partial rates to a charged \Kstar and a \Dz \CP-eigenstate divided by the 
\B \to \Dz \Kstar decay rate,
are sensitive to the
angle $\gamma$ 
of the CKM unitarity triangle. We measure:
\begin{eqnarray}
\Acpp &=& -0.09 \pm 0.20 (stat.) \pm 0.06 (syst.) \nonumber\\
\Acpm &=& -0.33 \pm 0.34 (stat.) \pm 0.10 (syst.) \ (+0.15 \pm 0.10) \cdot (\Acpm - \Acpp) \nonumber\\
\Rcpp &=& +1.77 \pm 0.37 (stat.) \pm 0.12 (syst.) \nonumber\\
\Rcpm &=& +0.76 \pm 0.29 (stat.) \pm 0.06 (syst.) \ ^{-\ 0.04}_{-\ 0.14} \nonumber
\end{eqnarray}
The third uncertainty quoted for the \cpm measurements reflects possible 
interference effects in the final states with $\phi$ and $\omega$ resonances. All results are preliminary.
\vfill
\begin{center}

Submitted to the 32$^{\rm nd}$ International Conference on High-Energy Physics, ICHEP 04,\\
16 August---22 August 2004, Beijing, China

\end{center}

\vspace{1.0cm}
\begin{center}
{\em Stanford Linear Accelerator Center, Stanford University,
Stanford, CA 94309} \\ \vspace{0.1cm}\hrule\vspace{0.1cm}
Work supported in part by Department of Energy contract DE-AC03-76SF00515.
\end{center}

\newpage

\begin{center}
\small

The \babar\ Collaboration,
\bigskip

%
B.~Aubert,
R.~Barate,
D.~Boutigny,
F.~Couderc,
J.-M.~Gaillard,
A.~Hicheur,
Y.~Karyotakis,
J.~P.~Lees,
V.~Tisserand,
A.~Zghiche
\inst{Laboratoire de Physique des Particules, F-74941 Annecy-le-Vieux, France }
% ...

\end{center}\newpage

\section{INTRODUCTION}
\label{sec:Introduction}
The measurement of \CP violation in \B meson decays offers the means to over-constrain the
unitarity triangle. A theoretically clean measurement of the angle 
$\gamma$=arg($-V_{ud}V^*_{ub}/V_{cd}V^*_{cb}$) is provided by
the \Bm \ra $D^{(*)0} \ K^{(*)-}$~\cite{bib:chargeconjimplied} decay
channels in which the 
favoured $b \ra c \ubar s$ and suppressed $b \ra u \cbar s$ penguin-less processes interfere~\cite{bib:theory}.
In this paper we implement the method proposed by Gronau, Wyler and London by looking at the interference between
\Bm \ra \Dz \Kstarm\ and \Bm \ra \Dzb \Kstarm\ when both \Dz and \Dzb\ decay
to a \CP\ eigenstate. \par
We define~\cite{bib:gronau}:
\begin{eqnarray}
\Acppm &=& \frac{\Gamma(\Bm\to D^{0}_{\CP\pm} \Kstarm) - \Gamma(\Bp\to D^{0}_{\CP\pm} \Kstarp)}
                        {\Gamma(\Bm\to D^{0}_{\CP\pm} \Kstarm) + \Gamma(\Bp\to D^{0}_{\CP\pm} \Kstarp)}, \\
\Rcppm &=& \frac{\Gamma(\Bm\to D^{0}_{\CP\pm} \Kstarm) + \Gamma(\Bp\to D^{0}_{\CP\pm} \Kstarp)}
                                               {\Gamma(\Bm\to\Dz\Kstarm)}.
\label{eq:acprcp}
\end{eqnarray}
\Acppm\ are the direct \CP-violating asymmetries associated to each \Dz \CP eigenstate. 
\Rcppm\ is twice the ratio of the charge-averaged
branching fraction of the \B \ra \dzcppm \Kstar decays to that of \Bm \ra \Dz \Kstarm . 
The latter branching fraction has been measured previously by \babar~\cite{bib:babarBR}, 
CLEO~\cite{bib:cleoBR} and Belle~\cite{bib:belleBR}. These
measurements average to $ (5.7 \pm 0.6)\cdot 10^{-4}$. 
Both \Acppm\ and \Rcppm\ carry \CP-violating
information. Neglecting \Dz -\Dzb\ mixing:
\begin{eqnarray}
\Rcppm	                &=& {1 \pm 2 r_{\B} \cos \delta \cos \gamma + r_{\B}^{2}}
\label{eq:cpinfo-rcp}
\end{eqnarray}
\begin{eqnarray}
\Acppm \times \Rcppm	&=& \pm \ 2 r_{\B} \sin \delta \sin \gamma
\label{eq:cpinfo}
\end{eqnarray}
where $\delta$ is the \CP-conserving strong phase phase difference between the \Bm \ra \Dz \Kstarm\ (favoured \B decay) 
and \Bm \ra \Dzb \Kstarm\ (suppressed) amplitudes, $\gamma$ is the \CP-violating weak phase difference in the same processes
and  $r_{\B} \simeq 0.1$-$0.3$ is the magnitude of the ratio of the amplitudes~\cite{bib:gronau}.

\section{THE \babar\ DETECTOR AND DATASET}
\label{sec:babar}
We look for \Bm \to \Dcp \Kstarm decays with the data collected near the \FourS\ resonance with the \babar\ detector
at the \pep2\ storage ring between October 1999 and May  2004. We accumulated an integrated luminosity of 205 \invfb\ on 
the peak (227 million \BB\ pairs) and 16 \invfb\ 40~\mev below the resonance.
The \babar\ detector is described elsewhere~\cite{bib:babar}. We focus on the components which are relevant to this
analysis. Charged-particle trajectories are measured by a five-layer double sided silicon vertex tracker (SVT) and a 40-layer
drift chamber (DCH) immersed in a 1.5~T solenoidal magnetic field. Charged-particle identification is achieved by combining
the light measured in a ring-imaging Cherenkov device (DIRC) with the ionization energy loss (\dedx )
measured in the DCH and SVT. Photons are detected in a CsI(Tl) electromagnetic calorimeter (EMC) inside the coil.
We use GEANT4~\cite{bib:geant4} based software to simulate the detector response and account for the varying beam and
environmental conditions.
\section{ANALYSIS METHOD}
\label{sec:Analysis}
We reconstruct \Bm \to \Dz \Kstar(892)$^-$ decays. We select  \Kstarm\ candidates in
the \Kstarm \to \KS \pim\ mode and \Dz\ candidates in 8 decay channels. We optimize our event-selection criteria to minimize the
statistical error on the signal yield, determined for each channel using simulated signal and background events.
In order to maximize statistics and as we measure ratios (asymmetries and ratios of branching fractions), we use rather relaxed criteria
to define a track or a photon candidate. In general a track must originate from the interaction point within 1.5~\cm\ in the
transverse plane and 10~\cm\ along the beam. Very soft tracks, measured only by the SVT are also used. Loose criteria are employed to
separate charged pions and kaons. Pions (kaons) are identified
with more than 95\% (90\%) efficiency by an algorithm which rejects more
than 85\% (95\%) of the kaons (pions) in the geometrical acceptance of
the DIRC. Photon candidates are isolated energy deposits in the EMC with an energy above 30~\mev\ and
a shape consistent with that of a photon-induced shower. \par
\KS\ candidates are made from oppositely charged tracks assumed to be
pions. \KS\ used to make  \dzcppm\ (\Kstar ) candidates are mass
(mass and vertex) constrained. The pion pairs are selected according
to their reconstructed invariant mass, required to be within 13~\mevcc\ from the
known value~\cite{bib:pdg2004}.  For those used to search for a \Kstar we further require their flight direction
and length to be as expected for a \KS\ coming from the interaction point. 
The \KS\ candidate flight path and momentum have to make an acute angle. The \KS\ candidate flight length in the transverse
plane has to exceed its uncertainty by 3~standard deviations. \piz\ candidates combine pairs of photons with a total energy
greater than 200~\mev\ and an invariant mass between 115 and 150~\mevcc . A mass constraint fit is applied to the selected
\piz\ candidates. \Kstar\ candidates combine a \KS\ and a charged particle which are constrained to come from a single decay
vertex. We select \Kstar\ candidates which have an invariant mass within 75~\mevcc\ of the known value~\cite{bib:pdg2004}.
Finally, since the  \Kstar in \B \to \Dz \Kstar\ is polarized, we
require $|$cos~$ \theta_{\Kstar hel}| \geq 0.35$, where
$\theta_{\Kstar hel}$ is the angle in the \Kstar\ rest frame between the daughter pion and
the parent \B momenta. Here and in other occurrences the helicity distribution
discriminates well between a \B\ meson decay and an event from the \epem \to \qqbar\ continuum.\par
 We select \Dz in even \CP\ (\cpp) eigenstates (\KpKm\ , \pip\pim),  odd \CP\ (\cpm) eigenstates (\KS\piz, \KS$\phi$ and \KS$\omega$), 
and non-\CP (flavour) eigenstates (\Km\pip , \Km\pip\piz\ and \Km\pip\pip\pim ). Composite
particles included in the \cpm\  modes are vertex constrained. $\phi$ ($\omega$) candidates are constructed from \Kp\Km (\pip \pim \piz) particle combinations with an invariant mass required to be within 12 (20)~\mevcc\ of the known
values~\cite{bib:pdg2004}. We further select the $\omega$ candidates with requirements on two helicity angles. The normal
helicity angle, between the \Dz\ momentum in the rest frame of the $\omega$ and the normal to the plane containing all 3
decay pions, must have its cosine above 0.25 in magnitude. The Dalitz-helicity angle~\cite{bib:dalitzomega}, defined as the angle
between the momentum of one daughter pion in the $\omega$ rest frame and the direction of one of the other two pions in the
rest frame of those two pions, must have a cosine with a magnitude less than 0.9. \par
Except for the \KS \piz\ final state, all \Dz candidates are mass and vertex constrained. We select \Dz\ candidates with an
invariant mass differing from the known mass by less than 12~\mevcc\ for all channels except \KS \piz\ (30~\mevcc) and \KS $\omega$
(20~\mevcc). These limits are about 2 standard deviations of the nearly-Gaussian resolutions. \par
To suppress the background due to \epem \to \qqbar reactions, $q\in \{u,d,s,c\}$, we require
$|$cos~$ \theta_B| \leq 0.9$ as, in the \FourS\ rest frame, \B mesons are produced with a $\sin^2 \theta_B$ distribution of the
angle $\theta_B$ of their momentum with the beam axis. We also use global event shape variables which translate quantitatively
the fact that in the \FourS\ rest frame, \qqbar\ continuum events have a two-jet like topology whereas \BB\ events
are more spherical. We require $|$cos~$ \theta_{thrust}| \leq 0.9$ where $\theta_{thrust}$ is
the angle between the thrust axes of the \B candidate and the rest of the event (i.e. what is left after removing the 
tracks and clusters associated with the \B candidate).
We construct a Fisher discriminant~\cite{bib:fisher} from cos~$\theta_{thrust}$ and Legendre monomials~\cite{bib:muriel}
describing the energy flow in the rest of the event. \par
We identify final \B candidates using two [nearly] independant kinematic variables: the beam-energy-substituted mass
$\mes=\sqrt{(s/2+{\bf p_0 \cdot p_B})^2/E_0^2-p_B^2}$ and the energy difference $\Delta E=E_B^*-\sqrt{s}/2$, where the
subscripts 0 and $B$ refer to the \epem-beam-system and the \B candidate respectively; $s$ is the square of the
center-of-mass (CM) energy and the asterisk labels the CM frame. For all signal modes, the \mes\ distributions are described by the same
Gaussian function $\mathcal{G}$ centered at the \B mass with a 2.6~\mevcc\ resolution (except \KS\piz which has a 2.8~\mevcc resolution).
The $\Delta E$ distribution depends more strongly on the signal mode.
\de\ is centered on zero for signal with a resolution of 11--13~\mev\ for all channels except 
\KS\piz\ for which the resolution is asymmetric and about 30~\mev. 
We define a signal region $|\Delta E| < 25$~\mev\ for all modes except \KS\piz which uses $<50$~\mev.
\par
For those events consistent with more than one \B\ candidate --- this occurs in $<25$\% of selected events depending on the \Dz mode --- 
we choose that with the smallest $\chi^2$ formed from the differences of the measured and true \Dz and \Kstar\
masses scaled by the mass spread which includes the resolution and, for the \Kstar, the natural width. Simulations prove
that no bias is introduced by this choice and the correct candidate is picked at least 82\% of the time. \par
The total reconstruction efficiencies\footnote{The branching fractions of the unstable daughter particles are not included.}
 after corrections, according to simulation of signal events, are:
12.5\% and 12.9\% for the \cpp modes \Dz \to \KpKm  and \pip\pim;
5.0\%, 10.2\% and 2.3\%  for the \cpm modes \Dz \to \KS\piz , \KS$\phi$ and \KS$\omega$;
13.5\%, 5.2\% and 8.0\% for the \Dz \to \Km\pip, \Km\pip\piz and \Km\pip\pip\pim\ non-\CP modes. \par

\section{PEAKING BACKGROUNDS}
\label{sec:Backgrounds}
To study \BB\ backgrounds we look in {\it sideband} regions away from the signal region. We define a \de\ sideband
in the  interval $-100 \leq \de \leq -60 $ and $60 \leq \de \leq 200 \mev $ for all modes
except \Dz \to \KS \piz\ for which the inside limit is $\pm95$ rather than 60~\mev. 
The lower limit ($-100$~\mevcc ) is chosen to avoid selecting much of the background coming from \Dstarz \to \Dz $\piz/\gamma$ decays.
In this \de\ sideband we see no significant evidence of a peaking background which could leak into the signal region.
We define a second control region in the $m_{\Dz}$ sideband away from the \Dz mass peak. This provides sensitivity to
background sources which mimic signal both in \de\ and \mes : the (doubly) peaking background. This pollution comes from
either charmed or charmless \B meson decays which do not contain a true \Dz. As many of the possible contributions
to this background are not well known, we attempt to measure this contribution by including the $m_{\Dz}$ sideband in the fit.
Relevant plots are shown on the third row of~\figref{asymmetryfit}.\par
A notable background for the \Bm \to (\Dz \to \pip\pim) (\Kstarm \to \KS \pim ) mode is the dacay 
$\Bm\to(\KS \pip\pim)_{\Dz}\pim$ which contains 
the same final state as the signal but has a 600 times higher branching fraction. 
We therefore explicitly veto any selected \B-candidate containing a (\KS\pip\pim) combination within 25~\mevcc~of the \Dz~mass.

\section{YIELD AND \boldmath{\CP} FIT}
\label{sec:fit}
An unbinned extended maximum likelihood fit to \mes distributions in the range $5.2\leq\mes\leq 5.3$~\gevcc 
is used to determine yields and \CP-violating quantities \Acp\ and \Rcp.
We use a {\it universal} Gaussian function $\mathcal{G}$ to describe the signal shape for all modes considered. The 
combinatorial background in the \mes\ distribution is modeled using a threshold function $\mathcal{A}{\it rg}$ 
first considered by the Argus collaboration~\cite{bib:argus}. Its shape is governed by one parameter $\xi$ that is
left to float in the fit and a second, E$_{MAX} = \sqrt{s}/2$, which is fixed at 5.2910~\gevcc.
We fit simultaneously \mes distributions of nine samples (1) the non-\CP , (2) \cpp\
and (3) \cpm \Dz modes in the ({\it i}) signal region, ({\it ii}) the
$m_{\Dz}$ sideband and ({\it iii}) the \DeltaE sideband.
We fit three types of probability density functions (\pdf ) combining $\mathcal{G}$(\mes ; $x_j$) and
$\mathcal{A}{\it rg}$(\mes ;~$\xi$) weighted by the unknown event
yields. We call $x_j$, the mean and standard deviation of $\mathcal{G}$ and
$\xi$ the shape parameters of the \pdf . In the \DeltaE sideband, we fit: 
$\mathcal{A}{\it rg}$. In the $m_{\Dz}$ sideband we fit: $ a_{sb} \cdot
\mathcal{A}{\it rg}$ + $b_{sb} \cdot \mathcal{G}$.  where $\mathcal{G}$
accounts for the doubly peaking \B-decays. Finally, we fit: 
$a \cdot \mathcal{A}{\it rg}+ b \cdot \mathcal{G}+ c \cdot \mathcal{G}$ in the signal region.
Here $b$ is scaled from $b_{sb}$ according to the $m_{\Dz}$ sideband
and signal window widths and is fixed; $c$ is the \Bpm\to\Dz\Kstarpm
signal. The motivation to perform the fit just described is as
follows: the non-\CP mode sample with relatively high statistics helps constrain the
\pdf\ shape for the low statistics  \CP mode distributions.
The \DeltaE sideband sample helps define the Argus background shape.

The values of $\xi$ obtained for each data sample were found to be consistent with each other, but subject to large uncertainties. We have therefore constrained $\xi$ to take the same value for all data samples in the fit.
We assume that the \B decays found in the $m_{\Dz}$
sideband have the same final states as the signal and
that we can therefore fit the same Gaussian to the signal and doubly
peaking \B background.\par

Furthermore, the \CP samples (in the signal region) are split into the \Bm and 
\Bp sub-samples. The likelihood function is written so as
to directly extract the four physical quantities, ${\cal A}_{+}$, ${\cal A}_{-}$, ${\cal R}_{+}$ and ${\cal R}_{-}$.
The doubly peaking \B-background is assumed to $not$ violate \CP and so
is split equally between the \Bm and \Bp\ sub-samples. 
This assumption is considered further when we discuss the systematic uncertainties.
Fig.~\ref{fig:asymmetryfit} shows graphically the results of the fit.
The nominal fit results are shown in table~\ref{tab:nominalfitresult}.\par

\Figure{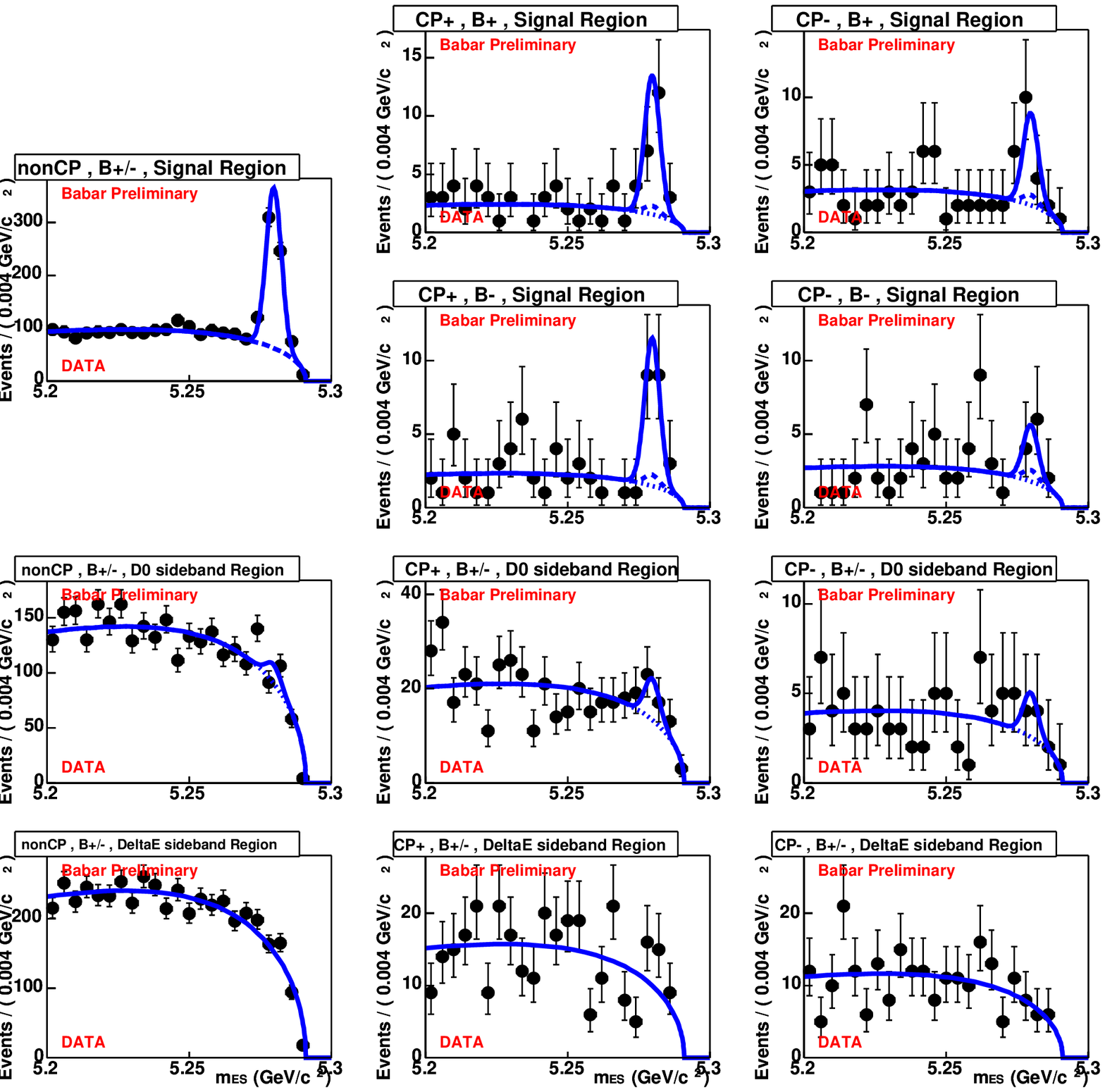}{
The simultaneous fit. The \mes distributions are shown for the non-\CP modes in the first column, \cpp\  modes in the second column and \cpm\ 
in the third. The fits in the signal region are shown in the top two rows, fits of the peaking background in the $m_{\Dz}$ sideband
are in the third row and the fits of the background in the \de sideband are in the fourth. For the \CP modes in the signal region, \Bp and \Bm distributions
are shown separately in the first and second rows.
}{17cm}

\begin{table}[htbp]
\begin{center}
\begin{tabular}{|l||r|r|r|}
\hline \hline
      & non-\CP     & \cpp\        & \cpm\ \\
\hline 
Yield     & 498 $\pm$ 29   & 34.4 $\pm$ 6.9     & 15.1 $\pm$ 5.8 \\
$N_{peak}$&10.9 $\pm$ 6.6  &$\ $ 2.5 $\pm$ 1.3  &$\ $2.4 $\pm$ 2.2 \\
\Acp  &                    &$-0.09 \pm$ 0.20    &$ -0.33 \pm$ 0.34\\
\Rcp  &                    &$\  1.77 \pm$ 0.37  &$\ $0.76 $\pm$ 0.29\\
\hline \hline
\end{tabular}
\caption{
\label{tab:nominalfitresult}
Results from the fits. For each \Dz\ mode class, we give the event
yield, the peaking background contribution, \Acp and \Rcp . 
The uncertainties are statistical only.}
\end{center}
\end{table}

The fit stability and accuracy have been studied by performing 1000 simulated
experiments at the optimum found by the fit in the parameter space.  
The value of the likelihood function in the 
fit falls well within the range observed with the simulated experiments.\par

The yields we observe for the non-\CP modes are compatible with our previous measurements of the \Bm \to \Dz \Kstarm\ branching ratio~\cite{bib:babarBR}. 
The statistical significance of the \cpp (\cpm) yields are 6.8~(2.9) standard deviations.
It should be noted that the measured \Acpm\ could differ from the
physical quantity, as background under the $\phi$ and $\omega$ resonances with
odd and even \CP could be present.
These possible interference effects are accounted for in the systematics.

\section{SYSTEMATIC STUDIES}
\label{sec:Systematics}
\subsection{\boldmath{\CP}-asymmetries}
Although most systematic errors cancel in the asymmetries, three effects must be considered. 
An asymmetry inherent to the detector or analysis may exist. After running the analysis code on a high statistics
\Bm\to\Dz\pim sample (with \Kstar cuts removed), the final selection shows an asymmetry of ($-1.9\pm0.8$)\%.
We quote a systematic uncertainty of $\pm$ 2.7\%.\par
The second substantial systematic effect is that due to a possible \CP-asymmetry in the peaking background.
Although there is no physics ansatz that requires the peaking background to be asymmetric, it cannot be excluded. 
To get an estimate, we remark that if there is an asymmetry $\mathcal{A}_{peak}$, a systematic error on \Acp\ is given
by $\mathcal{A}_{peak} \times \frac{b}{c}$, where $b$ is the contribution of the peaking background and $c$ the signal yield. 
Assuming conservatively $\mathcal{A}_{peak} \leq $ 50~\%, we obtain systematic errors 
of $\pm$~3.5~\% and $\pm$~8.0~\% on \Acpp\ and \Acpm respectively.\par 
Finally a systematic correction must be applied to \Acpm which stems from the background under the $\phi$ and $\omega$ resonances. 
Looking at the \Dz \to \Km\Kp\KS Dalitz plot~\cite{bib:antimo} we see
a contribution of the a$_0(980)$ under the $\phi$. 
The a$_0(980)$ is a $0^{++}$ state which induces a \cpp pollution. 
We crudely estimate a bias on \Acpm with a linear dependance on the
difference of the measured asymmetries, $\delta A_{\Acpm}=0.15 \pm 0.10 \cdot (\Acpm - \Acpp)$ 
which we quote as a separate systematic uncertainty.

\subsection{\Rcp}
\Rcp\ are ratios of rates of processes differing by the final state of
the \Dz . We have to consider the uncertainties affecting the
selection algorithms for the differing $D$ channels. 
They bring small correction factors which account for the difference
between the actual detector response and the simulation model. The main effects
stem from the approximate modeling of the tracking efficiency (1.2~\%
per track), the \KS\ reconstruction efficiency  for \cpm modes of
the \Dz\ (2.0~\% per \KS ), the \piz\ reconstruction efficiency for
the \KS \piz\ channel (3~\%) and the efficiency and misidentification
probabilities from the particle identification (2~\% per track). 
A substantial effect is the uncertainty on the measured branching ratios~\cite{bib:pdg2004}. 
We find 4.8~\% and 7.5~\% for the systematic uncertainties on the selection of the \Dz to \cpp 
and \cpm channels and 4.0~\% for \Dz\ to non-\CP channels.  Altogether, we find 
systematic uncertainties on the relative efficiencies to be 6.2~\% and 8.5~\% on
\Rcpp\ and \Rcpm respectively. Here also we quote as a separate systematic uncertainty, 
a possible bias due to structure under the $\phi$ and $\omega$ resonances. We conservatively 
estimate this bias to lie between $-$18 and $-$5~\% using the \Dz \to \Km\Kp\KS Dalitz plot~\cite{bib:antimo}.

\section{PHYSICS RESULTS}
\label{sec:Physics}
We quote the final results:

\begin{eqnarray}
\Acpp &=& -0.09 \pm 0.20 (stat.) \pm 0.06 (syst.) \nonumber\\
\Acpm &=& -0.33 \pm 0.34 (stat.) \pm 0.10 (syst.) \ (+0.15 \pm 0.10) \cdot (\Acpm - \Acpp) \ (bias) \nonumber\\
\Rcpp &=& +1.77 \pm 0.37 (stat.) \pm 0.12 (syst.) \nonumber\\
\Rcpm &=& +0.76 \pm 0.29 (stat.) \pm 0.06 (syst.) \ ^{-\ 0.04}_{-\ 0.14} \ (bias)\nonumber
\end{eqnarray}

For the \cpm measurements, the third uncertainty reflects a possible bias due to 
interference effects in the final states with $\phi$ and $\omega$ resonances.

These can be compared with the preliminary results by the Belle collaboration~\cite{bib:belleBR} using 95.8 \BB\ pairs:
$\Acpp = -0.02 \pm 0.33 \pm 0.07 ; \Acpm = +0.19 \pm 0.50 \pm 0.04$. The results are in agreement at the one 
standard deviation level.  As expected our statistical errors are reduced. We quote much larger systematic 
uncertainties because we account for a possibly high \CP-asymmetry in
the background and for structures under the $\phi$ and $\omega$
resonances whereas reference~\cite{bib:belleBR} 
does not. \par
We use equation~(\ref{eq:cpinfo-rcp}) to derive $r_{\B}^2=0.23 \pm 0.24$. Although the central value is large (it translates to 
$r_{\B}$=0.47), a null value is within one-standard deviation.

\section{SUMMARY}
\label{sec:Summary}
In summary, we present preliminary observations of the decays of charged B mesons to a
\Kstar\ and a \Dz where the latter particle is seen in final states of even and odd \CP . We express the
results in terms of \Rcp\ and \Acp . With more statistics, these quantities will 
constrain  $r_{\B}$, the ratio of amplitudes defined in
equations~(\ref{eq:cpinfo-rcp} and~\ref{eq:cpinfo}) and the angle $\gamma$ of the unitary
triangle by application of the Gronau-London-Wyler method. 

\section{ACKNOWLEDGMENTS}
\label{sec:Acknowledgments}

We are grateful for the 
extraordinary contributions of our \pep2\ colleagues in
achieving the excellent luminosity and machine conditions
that have made this work possible.
The success of this project also relies critically on the 
expertise and dedication of the computing organizations that 
support \babar.
The collaborating institutions wish to thank 
SLAC for its support and the kind hospitality extended to them. 
This work is supported by the
US Department of Energy
and National Science Foundation, the
Natural Sciences and Engineering Research Council (Canada),
Institute of High Energy Physics (China), the
Commissariat \`a l'Energie Atomique and
Institut National de Physique Nucl\'eaire et de Physique des Particules
(France), the
Bundesministerium f\"ur Bildung und Forschung and
Deutsche Forschungsgemeinschaft
(Germany), the
Istituto Nazionale di Fisica Nucleare (Italy),
the Foundation for Fundamental Research on Matter (The Netherlands),
the Research Council of Norway, the
Ministry of Science and Technology of the Russian Federation, and the
Particle Physics and Astronomy Research Council (United Kingdom). 
Individuals have received support from 
CONACyT (Mexico),
the A. P. Sloan Foundation, 
the Research Corporation,
and the Alexander von Humboldt Foundation.


\begin{thebibliography}{99}
\bibitem{bib:chargeconjimplied}
Charge conjugate channels are implied throughout the paper.
\end{thebibliography}
\end{document}